\documentclass{aa}
\usepackage{txfonts}

\usepackage{natbib}
\usepackage{txfonts}
\usepackage{graphicx}  
\bibpunct{(}{)}{;}{a}{}{,}
\usepackage[english]{babel}
\usepackage{ulem}
\usepackage{array}
\usepackage{hyperref}
\usepackage{enumerate}

\newcommand{\Porb}{\mbox{$P_\mathrm{orb}$}}
\newcommand{\taud}{\mbox{$\tau_\mathrm{dec}$}}
\newcommand{\td}{\mbox{$t_\mathrm{dec}$}}
\newcommand{\tq}{\mbox{$t_\mathrm{quiesc}$}}
\newcommand{\tr}{\mbox{$t_\mathrm{rise}$}}
\newcommand{\Rmax}{\mbox{$R_\mathrm{d,max}$}}
\newcommand{\Rd}{\mbox{$R_\mathrm{d}$}}
\newcommand{\Rave}{\mbox{$\langle{R_\mathrm{d}}\rangle$}}
\newcommand{\Msun}{\mbox{$\mathrm{M_{\odot}}$}}

\newcommand{\Tc}{\mbox{$T_\mathrm{c}$}}
\newcommand{\Mtr}{\mbox{$\dot{M}_{\mathrm{tr}}$}}
\newcommand{\Moutb}{\mbox{$\dot{M}_{\mathrm{outb}}$}}
\newcommand{\Mdot}{\mbox{$\dot{M}$}}

\newcommand{\Mcrp}{\mbox{$\dot{M}^+_{\mathrm{crit}}$}}

\newcommand{\Macc}{\mbox{$\dot{M}_{\mathrm{accr}}$}}
\newcommand{\Maccmax}{\mbox{$\dot{M}_{\mathrm{accr,max}}$}}
\newcommand{\Maccum}{\mbox{$\dot{M}_{\mathrm{accum}}$}}
\newcommand{\ve}{\mbox{$\mathrm{v}$}}
\newcommand{\An}{\mbox{$A_{\mathrm{n}}$}}
\newcommand{\Tn}{\mbox{$T_{\mathrm{n}}$}}
\newcommand{\Vmax}{\mbox{$M_\mathrm{V,max}$}}
\newcommand{\Vmin}{\mbox{$M_\mathrm{V,min}$}}

\def\apgt{\ {\raise-.5ex\hbox{$\buildrel>\over\sim$}}\ }
\def\aplt{\ {\raise-.5ex\hbox{$\buildrel<\over\sim$}}\ }

\newcommand{\ah}{\mbox{$\alpha_{\mathrm{h}}$}}
\newcommand{\ac}{\mbox{$\alpha_{\mathrm{c}}$}}

\newcommand {\be} {\begin{equation}}
\newcommand {\ee} {\end{equation}}
\newcommand{\beqa}{\begin{eqnarray}}
\newcommand{\eqa}{\end{eqnarray}}
\newcommand{\bea}{\begin{eqnarray}}
\newcommand{\eea}{\end{eqnarray}}
\newcommand {\bc} {\begin{center}}
\newcommand {\ec} {\end{center}}

\def\spose#1{\hbox to 0pt{#1\hss}}
\def\simless{\mathrel{\spose{\lower 3pt\hbox{$\mathchar"218$}}
        \raise 2.0pt\hbox{$\mathchar"13C$}}}
\def\simgreat{\mathrel{\spose{\lower 3pt\hbox{$\mathchar"218$}}
        \raise 2.0pt\hbox{$\mathchar"13E$}}}
\def\lta{\mathrel{\spose{\lower 3pt\hbox{$\mathchar"218$}}
        \raise 2.0pt\hbox{$\mathchar"13C$}}}
\def\gta{\mathrel{\spose{\lower 3pt\hbox{$\mathchar"218$}}
        \raise 2.0pt\hbox{$\mathchar"13E$}}}

\begin{document}

\title{The viscosity parameter $\alpha$ and the properties of accretion disc outbursts in close binaries}

\author{Iwona Kotko\inst{1}
\and
              Jean-Pierre Lasota\inst{1,2}
}
\offprints{Iwona.Kotko@uj.edu.pl}

\authorrunning{I. Kotko \& J.-P. Lasota}
\titlerunning{The viscosity parameter $\alpha$}

\institute{Astronomical Observatory, Jagellonian University, ul. Orla 171, 30-244 Krak\'ow, Poland
             \and
               Institut d'Astrophysique de Paris, UMR 7095 CNRS, UPMC Univ Paris 06, 98bis Bd Arago, 75014 Paris, France
              }
              \date{Received /
Accepted }              
\abstract{The physical mechanisms driving angular momentum transport in accretion discs are still unknown. Although it is generally accepted that, in hot discs, the turbulence triggered by the magneto-rotational instability is at the origin of the accretion process in Keplerian discs, it has been found that the values of the stress-to-pressure ratio (the $\alpha$  ``viscosity" parameter) deduced from observations of outbursting discs are an order of magnitude higher than those obtained in numerical simulations. }
{We test the conclusion about the observation--deduced value of $\alpha$ using a new set of data and comparing the results with model outbursts.}
{We analyse a set of observations of dwarf-nova and AM CVn star outbursts and from the measured decay times determine the hot-disc viscosity parameter $\ah$. We determine if and how this method is model dependent. From the dwarf-nova disc instability model we determine an amplitude vs recurrence-time relation and compare it to the empirical Kukarkin-Parenago relation between the same, but observed, quantities.}
{We found that all methods we tried, including the one based on the amplitude vs recurrence-time relation, imply $\ah \sim 0.1 - 0.2$ and exclude values an order of magnitude lower.}
{The serious discrepancy between the observed and the MRI--calculated values of the accretion disc viscosity parameter $\alpha$ is therefore real since there can be no doubt about the validity of the values deduced from observations of disc outbursts.}
\maketitle

\section{Introduction}
\label{intro}

Since the very beginning of the accretion-disc theory, the mechanism of angular momentum transport through the disc has been a matter of debate, and despite strenuous efforts of many researchers it remains an open problem because the results of numerical simulations do not match observations  when the respective values of the viscosity parameter $\alpha$ \citep{SS73} are compared \citep[][; see, however, Sorathia et al. 2012]{kingetal07}.  Magneto-rotational instability \citep[hereafter MRI;][]{Balbus98} simulations result in average $\alpha$ values of the order of $0.01$, whereas the best studied case of dwarf-nova outbursts unambiguously  provide values that are an order of magnitude higher \citep[][hereafter S99]{Smak99}.

In addition, from simulations of dwarf nova (DN) eruptions it is clear that $\alpha$ must vary during outbursts. The outburst amplitudes can be reproduced only if $\alpha_c$ in the quiescent disc is four to ten times smaller than $\alpha_h$ in the hot, outbursting state \citep[see e.g.][]{Smak84a,MMH84}. This holds regardless of the disc chemical composition \citep[however, the required $\alpha$ jump is rather 2 -- 6 for helium-dominated discs, see][]{KLDH1}. Since observations impose $\ah \approx 0.1- 0.3$, the cold disc \ac\, should be $\sim 0.01$. This is inconsistent with the MRI simulations that result in $\ah\sim 0.01$ \citep[see e.g.][]{Hirose09} and do not predict $\alpha$ jumps \citep[][obtain $\alpha$ jumps but with timescales too short to correspond to the dwarf-nova case.]{sorathiaetal12} In the present article we revisit the problem of \ah\, determination using a new set of data and methods that are somewhat different from those used by S99. We confirm his general conclusions about the value of $\ah$.

The values of the viscosity parameter $\alpha$ deduced from observations of dwarf-nova decay from outburst maximum are almost model-independent. Basically one only assumes  that the decay time reflects the viscous character of this process. 

On the other hand, the phenomenological relation between the outburst amplitude and the recurrence time \citep[the so-called Kukarkin-Parenago relation, hereafter K-P;][]{KP34,Warner03} obviously reflects some properties of the outbursts and should be derivable from the model supposed to be describing dwarf-nova outbursts, i.e. from the {\sl disc instability model} \citep[hereafter DIM; see][for a review]{L01}.  Using the DIM we derive an amplitude -- recurrence-time relation that compares reasonably well with the K-P relation, especially considering the large scatter of the observational data. We find that this K-P type relation also implies that $\ah$ cannot be of the order of 0.01 but must be roughly ten times larger than this value.

In Section \ref{sect:DIM} we briefly describe those aspects of the DIM that are required for understanding dwarf nova outbursts and the quantities that characterize them. The estimate of \ah \ from the empirical relation and from the analytical derivation of the outburst decay time are presented in Section \ref{sect:visc}. Section \ref{sect:KP} deals with the details of the analytical derivation of the Kukarkin-Parenago relation and its consequences for the parameter \ah. Conclusions are presented in Section \ref{sect:DC}.

\section{The dwarf nova outburst cycle in the DIM description}
\label{sect:DIM}

Outbursts of DN can be divided into two types:  normal outbursts and superoutbursts. As ``normal" one defines outbursts that are narrow, i.e. lasting no longer than about ten days, and that have amplitudes lower than $\sim 6$ mag and during which no superhumps (low-amplitude brightness variations) have been observed. Superhumps are the identification mark of the superoutbursts, which are also brighter and last longer than normal outbursts. In what follows ``outbursts" only refers to normal DN outbursts. DN showing only normal outbursts  are classified as U~Gem-type, whereas binaries exhibiting both normal and super outbursts belong to the SU UMa type.

In many DN  of U~Gem-type (e.g. in  \object{SS~Cyg} and \object{U~Gem}), a bimodality in the normal outbursts widths has been observed. 
The ``wide" outbursts are $\sim 0.2$ mag brighter and evidently longer than the ``narrow" outbursts. It has not been defined how much longer the outburst should be to be recognized as ``wide", but the difference is clear when looking at a specific light-curve. The wide outbursts should not be confused with superoutbursts, the former being normal DN outbursts with no superhumps detected, which in turn are the characteristic feature of the latter. According to the DIM, wide outbursts correspond to outside-in heating front propagation, while narrow outbursts result from the opposite sense of motion of these fronts \citep{Smak84a}.

During quiescence a dwarf nova disc accumulates matter until somewhere the temperature crosses the value critical to the onset of the thermal instability. A heating front starts propagating into the low-temperature regions, leaving behind ionized matter. If the radius where it started is close to the inner disc edge the front will propagate in the outward direction and the outburst will be of the inside-out type. If the front travels inwards from the outer edge, the outburst is an outside-in outburst. 

In the post-front hot regions of the disc the angular momentum transport is defined by \ah \ and the accretion rate \Macc \ rises. The enhanced efficiency of the outward angular momentum transport in the hot disc (\ah$>$\ac) causes mass, which during the quiescence had gathered in the outer parts of the disc, to  diffuse inwards at a high rate changing the surface-density $\Sigma$ profile from $\Sigma\sim R$ to $\Sigma\sim R^{-3/4}$. Once the heating front arrives at the outer (inner) disc edge,  matter across the whole disc becomes ionized, and \Macc \ everywhere in the disc becomes roughly constant and higher than the mass-transfer rate from the secondary (See Fig. \ref{fig:Alfa_KP_cycle}). In this phase the outburst is at its maximum, the disc is hot and the \citet{SS73} solution is a good approximation of such a quasi-stationary configuration. 

Because of ongoing accretion onto the central body the accretion rate (and density) decrease at a characteristic viscous timescale. Because the hot-disc critical surface density $\Sigma_{\rm crit}^+(R)\sim R^{1.11}$ (see Appendix \ref{App}),  $\Sigma$ on the outer edge easily falls below its critical  value and a cooling front starts propagating inwards (see Fig. \ref{fig:fronty}). The decay from outburst maximum can be seen as due to viscous emptying of a hot disc with a shrinking outer radius. 

Once the disc has lost matter accumulated during quiescence the refilling process restarts.  Strictly speaking the disc is filling up already during the outburst what allows the existence of outburst cycles with practically no quiescence.

\section{The value of the viscosity $\alpha$ parameter in the hot accretion discs}
\label{sect:visc}

As mentioned above the decay timescale is determined by the viscous timescale $t_{\rm visc}$ in the hot disc. Since $t_{\rm visc}\sim R^2/\nu$, assuming that the kinematical viscosity coefficient $\nu \sim \alpha c_s H$, where $c_s$ is the sound speed and $H$ the disc semi-thickness, determining the decay time allows estimation of $\alpha$, assuming one can estimate the disc radius (e.g. from the binary's orbital period).

S99 obtained an observational relation between the outburst decay rate \taud (or alternatively the outburst width $W$) and the orbital period \Porb: $\td(\Porb)$ and $W(\Porb)$. After comparing it with the relations of the same type found from the fits to data from the numerical models, he concluded that the best agreement  between observation and models is obtained for \ah$=0.2$. Below, we determine the viscosity parameter using a different DN sample and different numerical models.

\subsection{$\alpha$ from the decay rate vs. the disc radius relation}
\label{sect:tdecRmax}

To be certain that the result is independent of the choice of the DN sample for deriving the $\taud-\Porb$ relation, we used a set of observational data  different from S99, who used the data from \citet{VP83} and \citet{Warner03} that include all types of DN: the U~Gem, SU~UMa, Z~Cam stars, while our sample contains systems from \citet{Ak02}, together with an additional seven U~Gem-type DN for which we measured the decay rates using light-curves from the AFOEV database. Among the twenty one systems that we took into account nine have been included also in S99. 
 
Since it has been shown that the basic DIM (with no mass transfer rate enhancement or the additional sources of disc heating) is only able to reproduce normal outbursts  \citep[see][and references therein]{L01,KLDH1}, we find it reasonable to start our analysis with U~Gem-type systems where only normal outbursts are present\footnote{In U~Gem itself, one superoutburst has been observed \citep[e.g.][]{Mason88,SmakWaagen}} to ensure that the comparison between $\taud$ measured in the models and in the real light-curves is consistently defined. Then, to test to what extent other DN types and the disc chemical composition influence the correlation between \taud \ and \Porb, normal outbursts of  SU~UMa stars and one AM~CVn system (\object{PTF1J0719}) have been added to the sample. 

\begin{figure}[h]
\begin{center}
\includegraphics[scale=0.85]{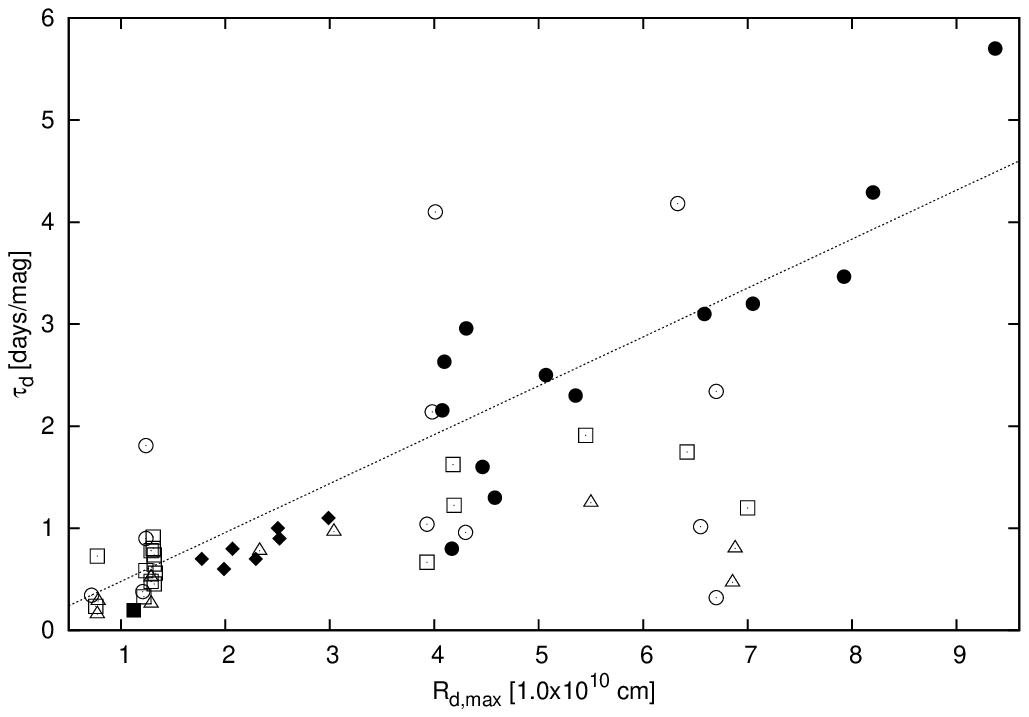}
\caption{\footnotesize The \taud-\Rmax \ relation. \textit{Filled symbols}: U~Gem-type systems (\textit{circles}), normal outbursts of SU~UMa-type systems (\textit{diamonds}), and PTF1J0719 (\textit{square}). \textit{Open symbols}: models with $\ah=0.1$ (\textit{circles}), $\ah=0.2$ (\textit{squares}), $\ah=0.3$ (\textit{triangles}). \textit{The dotted line} - linear fit to the observational data (filled symbols) in the form  $\taud=A_1\Rmax$, with $A_1=0.48$. (For the sake of clarity models with $\ah=0.05$ were not plotted.)}
\label{Fig:tdec}
\end{center}
\end{figure}

The systems considered  are listed in Table \ref{Tab:UGem} and marked in Fig. \ref{Fig:tdec}. The decay rates of U~Gem-type binaries (rows $1-8$) and the decay rates for the normal outbursts in SU~UMa-type binaries (rows $14-20$) are taken from \citet{Ak02}.  In rows $9-13$ are $7$ U~Gem-type binaries not included in \citet{Ak02}. For each of them we have measured \taud \ as the time it takes the system brightness to decline by $\sim 1$ mag starting from the level $1$ mag below the maximum. The last row stands for the only AM~CVn-type star for which the existence of the normal outburst have been confirmed \citep{Levitan11}.
\renewcommand{\arraystretch}{1.3}
\begin{table*}\footnotesize
\caption{Decay-from-outburst properties of selected dwarf novae.}
\begin{centering}
\begin{tabular}{|c|c|c|c|c|c|c|c|c|c|}
\hline 
 & System & \Porb & $M_1$ & $M_2$ & \Rmax & W & $\taud^o$ & $\taud^t_{0.2}$ \tabularnewline
 &              &  (hr)   & ($\Msun$) & ($\Msun$) & ($10^{10}$cm) & (d) & (d/mag) & (d/mag) \tabularnewline
\hline 
\hline 
$1$ & \object{BV Cen} & $14.67$ & $1.24$ & $1.1$ & $9.76$ & $20.9$ & $5.7$ & $5.47$ \tabularnewline
\hline 
$2$ & \object{AT Ara} & $9.01$ & $0.53$ & $0.42$ & $5.354$ & $4.1$ & $2.3$ & $1.98$ \tabularnewline
\hline 
$3$ & \object{RU Peg} & $8.99$ & $1.21$ & $0.94$ & $7.05$ & $7.2$ & $3.2$ & $2.75$ \tabularnewline
\hline 
$4$ & \object{MU Cen} & $8.21$ & $1.2$ & $0.99$ & $6.586$ & $7.9$ & $3.1$ & $4.15$ \tabularnewline
\hline 
$5$ & \object{SS Cyg} & $6.6$ & $0.81$ & $0.55$ & $5.069$ & $6.4$ & $2.5$ & $1.89$ \tabularnewline
\hline 
$6$ & \object{TWVir} & $4.38$ & $0.91$ & $0.4$ & $4.168$ & $3.7$ & $0.8$ & $1.82$ \tabularnewline
\hline 
$7$ & \object{SS Aur} & $4.33$ & $1.08$ & $0.39$ & $4.462$ & $4.3$ & $1.6$ & $1.48$ \tabularnewline
\hline 
$8$ & \object{U Gem} & $4.25$ & $1.2$ & $0.42$ & $4.58$ & $3.2$ & $1.3$ & $1.32$ \tabularnewline
\hline 
\hline
$9$ & \object{EY Cyg} & $11.02$ & $1.1$ & $0.49$ & $8.201$ & $14.3$ & $4.29$ & $2.83$ \tabularnewline
\hline 
$10$ & \object{DX And} & $10.57$ & $1.2$ & $0.8$ & $7.923$ & $12.5$ & $3.47$ & $2.90$ \tabularnewline
\hline 
$11$ & \object{EX Dra} & $5.04$ & $0.75$ & $0.56$ & $4.097$ & ? & $2.63$ & $2.01$ \tabularnewline
\hline 
$12$ & \object{BD Pav} & $4.30$ & $1.15$ & $0.73$ & $4.307$ & ? & $2.96$ & $1.54$ \tabularnewline
\hline  
$13$ & \object{IP Peg} & $3.797$ & $1.16$ & $0.55$ & $4.077$ & ? & $2.16$ & $1.75$ \tabularnewline
\hline 
\hline
$14$ & \object{CU Vel} & $1.884$ & $1.23$ & $0.15$ & $2.988$ & $4.5$ & $1.1$ & $1.05$ \tabularnewline
\hline 
$15$ & \object{WX Hyi} & $1.796$ & $0.9$ & $0.16$ & $2.519$ & $4.0$ & $0.9$ & $1.55$ \tabularnewline
\hline 
$16$ & \object{Z Cha} & $1.788$ & $0.84$ & $0.13$ & $2.503$ & $3.7$ & $1.0$ & $1.24$ \tabularnewline
\hline 
$17$ & \object{VW Hyi} & $1.783$ & $0.67$ & $0.11$ & $2.292$ & $4.0$ & $0.7$ & $0.90$ \tabularnewline
\hline 
$18$ & \object{OY Car} & $1.515$ & $0.64$ & $0.086$ & $2.07$ & $4.7$ & $0.8$ & $0.74$ \tabularnewline
\hline 
$19$ & \object{Ek TrA} & $1.509$ & $0.46$ & $0.09$ & $1.775$ & $3.0$ & $0.7$ & $0.50$ \tabularnewline
\hline 
$20$ & \object{SW UMa} & $1.364$ & $0.71$ & $0.1$ & $1.987$ & $7.6$ & $0.6$ & $0.49$ \tabularnewline
\hline 
\hline
$21$ & \object{PTF1J0719} & $0.446$ & $0.5$ & $0.05$ & $0.871$ & $1$ & $0.25$ & $0.79$ \tabularnewline
\hline 
\end{tabular}
\par\end{centering}
\centering{}
\tablefoot{{\footnotesize  $\Porb$ is the orbital period, $M_1$ and $M_2$ the primary and secondary masses respectively, $\Rmax$ the maximum disc radius, W the observed outburst width, $\taud^o$ the observed decay rate, and $\taud^t_{0.2}$ the theoretical (Eq. \ref{eq:tdec}) decay rate for $\ah=0.2$. \textit{Rows}: $1-8$: U~Gem stars from \citet{Ak02}, $9-13$: U~Gem stars with \taud \ calculated by authors from light-curves from AFOEV database, $14-20$: SU~UMa stars from \citet{Ak02}, $21$: AM~CVn star from \citet{Levitan11}; $\Porb$, $M_1$ and $M_2$ are taken from \citet{Ritter03}; \Rmax \ is calculated according to Eqs. (\ref{eq:Rd}) \& (\ref{eq:a}). }}
\label{Tab:UGem}
\end{table*}

First, we compared our fit to the \taud \, -- \Porb\,  relation with the one obtained by S99. The linear fit to our data of the form $\taud=C_{\tau}\Porb$ \citep[the ``Bailey relation";][]{Bailey75}  gives $C_{\tau}=0.37 \pm 0.03$ with dispersion $rms=0.698$ which compares nicely with the result of S99: $C_{\tau,{\rm S99}}=0.38 \pm 0.02$ with $rms=0.54$. For the more general case when $\taud=C_{\tau}^1\Porb^{\beta}$ we get $C_{\tau}^1=0.69 \pm 0.17$ and $\beta=0.66 \pm 0.14$ with dispersion $rms=0.64$, while S99 result is $C_{\tau,{\rm S99}}^1=0.61 \pm 0.07$ and $\beta_{\rm S99}=0.71 \pm 0.09$ with dispersion $rms=0.48$. Our results are similar to those of S99 within the uncertainty of the fit coefficients.

Next, we compared model decay times with those observed during dwarf-nova outbursts. Because models are calculated not for a given \Porb\, but for a given disc radius, we convert orbital periods of observed systems into the disc radii.

During outburst the outer disc radii in DN expand up to $\Rmax \sim 0.9R_{\rm L_1}$  \citep[see e.g.][]{Smak01}, where $R_{\rm L_1}$ is the radius of a primary Roche-lobe given by the \citet{egg83} formula:
\be
\frac{R_{\rm L_1}}{a}=\frac{0.49\,q^{2/3}}{0.6\,q^{2/3}+\log\left(1+q^{1/3}\right)}
\label{eq:Rd}
\ee
\be
a=3.5\times 10^{10} M_{2}^{1/3}(1+q)^{1/3}P_{\rm hr}^{2/3}\,{\rm cm}
\label{eq:a}
\ee
where $q=M_1/M_2$ and $P_{\rm hr}$ is the orbital period in hours.

The primary and the secondary masses ($M_1$ and $M_2$) for calculations of \Rmax \ from \Porb \ were taken from the latest version (November $2011$)  of \citet{Ritter03}. For PTF1J0719, $M_1$ and $M_2$ have been guessed according to what is expected for AM~CVn stars since no observational estimates have been suggested yet. The fit in the form $\taud=A_1\Rmax$ to all data from Table \ref{Tab:UGem} gives $A_1=0.48 \pm 0.02$ with dispersion $rms=0.56$.

It is interesting to check whether the $\taud-\Rmax$ relation is independent of the class of systems exhibiting normal outbursts. The \taud--\Rmax \ relation for U~Gem-type binaries is  linear to a good approximation.  For the linear fit one obtains $A_1=0.49 \pm 0.03$ with dispersion $rms=0.68$. The coefficients for the general relation $\taud=B_1\Rmax^{\gamma}$ are $B_1=0.27 \pm 0.12$ and $\gamma=1.32 \pm 0.22$ with $rms=0.66$. In this case, similar to what was noticed by S99, the $rms$ dispersions do not differ significantly for the linear and non-linear fits, moreover the errors of $B_1$ and $\gamma$ are rather large. Therefore we limit our further considerations to the simpler linear case.

According to the model, the outbursts appearing between superoutbursts of SU~UMa stars have the same origin as those in U~Gem-type binaries. As expected, their measured \taud \ marked on the \taud--\Rmax \ plane extrapolate the \taud--\Rmax \ relation for U~Gem to the regime of orbital periods shorter than $2\, {\rm hr}$. The coefficient $A_1$ of the linear fit for the sample, including the normal outbursts of U~Gem-type and SU~UMa-type DN is $A_1=0.48 \pm 0.03$ with dispersion $rms=0.57$. 

As discussed in \citet{KLDH1}, PTF1J0719 is the only system in the AM~CVn class of binaries where short outbursts can be firmly classified as ``normal" and  the system considered as a helium counterpart of an SU~UMa-type DN. With PTF1J0719 taken into account, the coefficient of the linear fit remains almost unchanged. 

We conclude that the $\taud-\Rmax$ relation is universal for normal outbursts of all classes of cataclysmic variables. 
To estimate the \ah \ parameter one needs to find the relevant \taud-\Rmax \ relation for  model light-curves calculated with different \ah \ and compare the result with observations. 
We chose four values of \ah: $0.05$, $0.1$, $0.2$,  and $0.3$, for each of them the set of models with different mass transfer rates \Mtr, primary masses $M_1$ and maximum disc radii \Rmax \ were calculated. The decay rates of the synthetic outbursts were measured in the same manner as in the observational case. 

To obtain the synthetic light-curves we used the code described in \citet{H98}, which differs from the code used by S99; for example the input parameters are different,\footnote{S99 uses $M_1$ and $M_2$ (which define the orbital period of a semi-detached binary system) as the input parameters  and the mean disc radius \Rave, while in the \citet{H98} code, the input is $M_1$ and \Rave. Smak defines only one value of $\alpha$, while we take as the input parameters both \ah \ and \ac.} and in our code the adaptive grid enables high resolution of the fronts. Such differences should not affect modelling of the decay phase of the outburst.

When measuring \taud \ of the outbursts for a wide range of model parameters, one has to pay attention to several problems:
\begin{enumerate}[(a)]
\label{criteria}

\item For large discs ($\Rmax>5.0\times 10^{10}\, {\rm cm}$) and high primary masses ($M_1>1\,\Msun$), so-called ``reflares" appear during the decline from maximum. They are an indication of the cooling and heating front reflections in the disc, where the surface density is close to its critical value \citep[for details see e.g.][]{Menou00,Dubus01}. In this case the model outburst cannot be considered as normal.

\item In large discs and for high values of \ah \ , inside-out heating fronts may not be able to propagate up to the outer disc edge because such values of the viscosity parameter decrease the value of $\Sigma_{\rm crit}^+$ (see the formulae in Appendix \ref{App}) and, with increasing $R$, increase the possibility of a cooling front forming right behind the heating front. In this case the cooling front will start at $R\ll \Rmax$ and the decay rate will not be connected with the actual radial extent of the disc. 

\item Models with the same parameters except for \ah \ have different stability limits. 
\end{enumerate}

The above-mentioned DIM properties have their reflection in the distribution of the model points in Fig. \ref{Fig:tdec}. For large \Rmax \ models with higher \ah \ tend to deviate more from the empirical \taud-\Rmax \ relation.

Linear fits to the $\taud=A_{\alpha}\Rmax$ relation obtained for the models with different \ah \ give 
\begin{enumerate}
\item $A_{0.05}=1.624 \pm 0.235$ for \ah$=0.05$,
\item $A_{0.1}=0.525 \pm 0.128$ for \ah$=0.1$,
\item $A_{0.2}=0.338 \pm 0.036$ for \ah$=0.2$,
\item $A_{0.3}=0.151 \pm 0.031$ for \ah$=0.3$.
\end{enumerate}
The coefficients $A_{\alpha}$ show a clear tendency to decrease when a higher \ah \ is set in the model.
The comparison with $A_1$ obtained from the fits to empirical data shows that \ah$\in [0.1,0.2]$, with no unambiguous preference for one of these values, thus confirming conclusions obtained by S99.

SU~UMa-type stars and their superoutbursts provide another piece of interesting information.
The decay from superoutburst may be divided  into at least two phases - the plateau and fast decay phases. According to the enhanced mass-transfer (EMT) model \citep[][]{KLDH1,Smak08,Smak09a,Smak09b,Smak09c,Smak09d}, during the plateau phase the slow decline of the system luminosity is caused by accretion-driven depletion of the excess matter provided by the enhanced mass transfer from the secondary. This phase ends when a cooling front forms, and so the following fast decline is caused by the mechanism producing normal outbursts.

Based on this, we measured \taud \ during the fast decay phase of SU~UMa superoutbursts and found that they are approximately the same as the $\taud$  measured for their normal outbursts. The same is true of superoutbursts and normal outbursts in the models calculated with the prescription for the $\Mtr$ enhancement given  in  \citet{H97}. The decay time in the fast decline phase was measured as the time interval between the time the system luminosity was $1$ mag below the start of the decline phase to the time when the system was $2$ mag below it.

This conclusion is very promising in the context of evaluating \ah \ in AM~CVn stars. As already mentioned, the normal outbursts in AM~CVn stars are rarely detected, and the outburst cycle is dominated by superoutbursts. However, with well observed, fast decay phases of the superoutbursts in AM~CVn stars, it will be possible to estimate \ah \ in helium-dominated discs more precisely. Unfortunately the currently available data are not of sufficient quality to permit such investigations.

\subsection{The outburst width -- orbital period \ relation}
\label{sect:WPorb}

\citet{VP83} showed that there exists a positive correlation between the outburst width $W$ and the orbital period but concluded that narrow and wide outbursts should be considered separately. 

To  consistently determine the outburst width in various systems, the magnitude level at which it is measured has to be defined. Following \citet{VP83} S99  defines $W$ as the time interval during which the system luminosity is above the level set at $2$ mag below the outburst maximum.

Using data from \citet{VP83}, S99 finds the coefficients for
\begin{itemize}\item the linear dependence in the form 
$W=C_{W,\rm S99}\Porb$: $C_{W,{\rm S99}}=1.39 \pm 0.06$;
\item the non-linear dependence in the form $W=C_{W,{\rm S99}}\Porb^{\beta_{\rm S99}}$: $C_{W,{\rm S99}}^1=2.01 \pm 0.29$ and $\beta_{\rm S99}=0.78\pm 0.11$. 
\end{itemize}
Since, as in S99, we find the linear fit to be of  superior quality, in what follows we do not use  the non-linear fitting formula. We also only use  narrow outbursts as ``generic" normal outbursts. The linear fit to our data (18 systems) gives  $C_W=0.99 \pm 0.12$, so the agreement with S99 is not as good as for the decay times.

The analogous procedure applied to different subsets of our data gives
\begin{itemize}
\item for U~Gem stars from \citet{Ak02} only: $C_W=0.79 \pm 0.07$;

\item for U~Gem stars from \citet{Ak02} complemented with our measurements: 
$C_W=0.90 \pm 0.10$. 

\end{itemize}
One concludes that $C_W$ depends on the choice of the DN sample. 

The measurements of the outbursts width are clearly more vulnerable to uncertainties that are connected with the precise determination of the outburst maximum and with usually sparser data coverage of the outburst rise in comparison to the outburst decline. Moreover, except for systems observed intensively for a long time (such as the  already mentioned SS~Cyg or U~Gem), straightforward assessment of which outbursts are narrow and which are wide may be problematic. Nevertheless, it is worth comparing observations with models as has been done in Section  \ref{sect:tdecRmax} for \taud. The width $W$ of synthetic outbursts has been defined in the same manner as in the observational case. 

When applying the linear dependence $W=C_{\alpha}\Rmax$ to the same set of models as in Section \ref{sect:tdecRmax}, one obtains
\begin{enumerate}
\item $C_{0.05}=3.222 \pm 1.132$ for \ah$=0.05$,
\item $C_{0.1}= 1.794 \pm 0.285$ for \ah$=0.1$,
\item $C_{0.2}=1.502 \pm 0.112$ for \ah$=0.2$,
\item $C_{0.3}=0.872 \pm 0.165$ for \ah$=0.3$.
\end{enumerate}
Comparison of the above listed $C_{\alpha}$'s with $C_1=1.496 \pm 0.135$ determined from the observed relation $W=C_1\Rmax$ again favours  \ah$\in [0.1,0.2]$ with even stronger indication of \ah$=0.2$.

\begin{figure}[h]
\begin{center}
\includegraphics[scale=0.85]{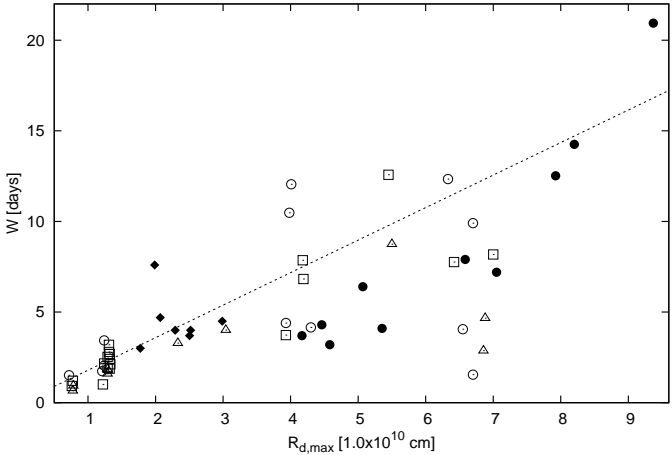}
\caption{\footnotesize The $W$-\Rmax \ relation. \textit{Filled symbols}: U~Gem-type systems (\textit{circles}), normal outbursts of SU~UMa-type systems (\textit{diamonds}), and PTF1J0719 (\textit{square}). \textit{Open symbols}: models with $\ah=0.1$ (\textit{circles}), $\ah=0.2$ (\textit{squares}), $\ah=0.3$ (\textit{triangles}). \textit{The dotted line} - linear fit to the observational data (filled symbols) with a coefficient $C_1=1.496$. (For the sake of clarity models with $\ah=0.05$ were not plotted.)}
\label{Fig:W}
\end{center}
\end{figure}

The model and observational data with fitted linear dependence between $W$ and \Rmax \ are presented in Fig. \ref{Fig:W}.
\begin{table*}
\caption{Decay rates for several solar composition disc models.}
\begin{centering}
\begin{tabular}{|c|c|c|c|c|c|c|c|}
\hline 
Model & $\alpha_{\mathrm{h}}$ & $M_{1}$ & $R_{\mathrm{d,max}}$ & $A_{\mathrm{n}}$ & $\taud^m$ & $\taud^t$ & $\taud^v$ \tabularnewline
            &                                      &    ($\Msun$)  &   ($10^{10}$\, cm)  & (mag)  & (d/mag)  & (d/mag) & (d/mag) \tabularnewline
\hline 
\hline 
\hline 
1. & $0.2$ & $1.0$ & $4.18$ & $6.3$ & $1.62$ & $1.17$ & $8.38$\tabularnewline
\hline 
2. & $0.2$ & $1.0$ & $1.29$ & $4.1$ & $0.78$ & $0.6$ & $7.17$\tabularnewline
\hline 
3. & $0.2$ & $0.6$ & $1.29$ & $5.6$ & $0.48$ & $0.42$ & $7.65$\tabularnewline
\hline 
4. & $0.2$ & $0.6$ & $0.76$ & $6.3$ & $0.23$ & $0.23$ & $2.75$\tabularnewline
\hline 
5. & $0.3$ & $1.3$ & $3.04$ & $3.3$ & $0.97$ & $1.13$ & $10.37$\tabularnewline
\hline 
6. & $0.3$ & $1.0$ & $1.29$ & $3.2$ & $0.52$ & $0.51$ & $6.1$\tabularnewline
\hline 
7. & $0.3$ & $1.0$ & $0.78$ & $3.7$ & $0.29$ & $0.27$ & $4.4$\tabularnewline
\hline 
8. & $0.3$ & $0.6$ & $0.77$ & $3.7$ & $0.16$ & $0.26$ & $3.14$\tabularnewline
\hline 
\end{tabular}
\par
\end{centering}
\tablefoot{$\taud^m$ is
measured from the model, $\taud^t$
calculated from Eq. (\ref{eq:tdec}), and
$\taud^v$  from Eq. (\ref{tvis}).}
\label{decaytimes}
\end{table*}
Even if $W$ is not determined well enough  to provide a firm value of \ah \ from the $W-\Rmax$ relation, the results obtained totally preclude $\ah\, \ll 0.1$.

\subsection{The decay time from the DIM}
\label{sect:tdec}

Since we have assumed that the decay from outburst's maximum is described by the standard DIM, it is worth checking how strongly model-dependent the results are concerning the value of the viscosity parameter. In principle, the setting is very simple: shrinking (from outside) hot--disc configurations decay through a set  a quasi-stationary solutions corresponding to monotonically diminishing accretion rates. This prompted S99 to estimate the decay time as:
\be
\td\sim\frac{\Rmax}{\ve_{\rm vis}},
\label{tvis}
\ee
where the viscous speed ${\ve_{\rm vis}}\sim \nu/R$.
However, S99 used a non-standard definition of the kinematic viscosity coefficient. To clarify this point we recall that
the idea behind the \citet{SS73} $\alpha$ ansatz  is that the component $\tau_{r\phi}$ of the stress-tensor should be proportional to the total pressure $P$ with the proportionality constant $\alpha$ describing the efficiency of the angular momentum transport due to turbulence in the disc: $\tau_{r\phi}=-\alpha P$, where $0<\alpha<1$. On the other hand, in a differentially rotating fluid the tangential stress is defined as $\tau_{r\phi}=\eta {d\Omega}/{dR}$, where $\eta$ is the dynamical viscosity coefficient, $R$ the radius, and $\Omega$ the angular speed. For a Keplerian disc, therefore, the kinematic viscosity coefficient $\nu=\eta/\rho$, where $\rho$ is the density, can be written as
\be
\nu=\frac{2}{3}\alpha\frac{c_s^2}{\Omega_K}\approx \frac{2}{3}\alpha c_s H
\label{eq:nu}
\ee
where $c_s$ is the sound speed,  $\Omega_K$ the Keplerian angular speed., and $H\approx c_s/\Omega_K$ the disc (semi) height-scale and not the disc's actual (e.g. photospheric) height $z_0$ as assumed in S99.  

Fortunately,  $\ah$  was not calculated in S99 directly from the analytical formula but obtained (as we did in the previous sections) from comparing the theoretical and empirical dependence between the outburst decay rate (or the outburst width) and the orbital period, so this article's conclusions concerning the value of the viscosity parameter in hot accretion discs remain valid. Nevertheless, it would be useful to derive an analytical formula for \taud, which would only depend on \ah \ and observables since it would allow dispensing with using numerically calculated models, especially because these must be selected according to criteria discussed in Sect. \ref{sect:tdecRmax}. In Table \ref{decaytimes} we compared decay rates calculated from numerical models ($\taud^m$) with the values ($\taud^v$) corresponding to Eq. (\ref{tvis}). (In the selected models fronts always propagate through the whole disc extent.) Clearly the rates obtained from Eq. (\ref{tvis}) are much too long and the approximation used was too crude.

Indeed, the decrease in luminosity after the outburst maximum is the effect of two mechanisms: (1) the depletion of the matter from the disc due to (viscous) accretion onto the central object, and (2) the propagation of the cooling front through the disc \citep[also a viscous process, for details see][]{Menou99,L01}. Taking the disc shrinking into account  lowers the decay time.

W assume that the decay time \td \ is the time it takes the system luminosity to drop from the maximum to the quiescence level and that it may be written approximately as
\be
\td\approx\frac{\Rmax}{|\ve_{\rm dec}|}
\ee
where (as before) \Rmax \ is the maximum disc radius and $\ve_{\rm dec}$ a decline velocity.

To find a simple formula we assumed that  $\ve_{\rm dec}$ is the  superposition of (a) the inward viscous velocity $\ve_{\rm visc}$ of matter in the hot part of the disc into which the cooling front propagates and (b) the outward velocity of the gas at the front $\ve_{F}$\footnote{We follow here the  \citet{Menou99} notation according to which $\ve_{F}$ is {\sl not} the front velocity but the gas velocity at the front. The ratio of the gas velocity at the cooling front to the cooling front velocity is typically $\sim 2$} \citep{Menou99} allowing the inward propagation of the front; i.e.  $|\ve_{\rm dec}|\approx|\ve_{\rm visc}+\ve_{F}|$. This crude approximation treating the two velocities as an average over space values gives quite good results, as seen in Table \ref{decaytimes}. This is because for most of the disc's extent \citep[in the ``asymptotic regime":][their Fig.~7]{Menou99},  $\ve_{F}$ is roughly constant.
Since $\ve_{\rm visc} \approx\nu/R$, from Eq.~(\ref{eq:nu}) one has
\be
\ve_{\rm visc}\approx\frac{2}{3}\ah c_s^2\frac{1}{\Omega_K R}
\label{eq:vvisc}
\ee
where $P$ is the total pressure and $c_s=\sqrt{P/\rho}$.

From numerical simulations, \citet{Menou99} find that $\ve_{F}\sim 1/7\ah\, c_s$. We confirmed this result for cooling front velocity in solar composition discs and found that it also applies to a helium-dominated disc. The final formula for \td \ is thus:
\be
\td \approx \frac{7\Rmax}{\ah c_s \left(1 +{14/3}\left(c_s/\ve_{K}({\Rmax})\right)\right)}.
\label{eq:tdec}
\ee
The speed of sound can be expressed in terms of the central temperature in the disc: $c_s=\sqrt{k\Tc  /m_H}$, where $k$ is the Boltzmann constant and $m_H$  the hydrogen molecular mass. (In the case of a helium disc it should be replaced by helium molecular mass). 
The numerical fit to the temperature at the cooling front found from models of solar-composition discs gives
\be
\Tc  \approx 4.7\times 10^4\,{\rm K},
\label{eq:fitTcH}
\ee
with no dependence on disc parameters. From Table \ref{decaytimes} one can see that Eq. (\ref{eq:tdec}) can give reliable estimates of of the viscosity parameter, certainly better than Eq. (\ref{tvis}).

For AM~CVn stars, e.g. PTF1J0719, the chemical composition (${\rm Y}=0.98\,\,{\rm Z}=0.02$) gives
\be
\Tc  \approx 1.1 \times 10^5\,{\rm K}
\label{eq:fitTcHe}
\ee
From Eqs. (\ref{eq:tdec}) and (\ref{eq:fitTcH}) (or (\ref{eq:fitTcHe})) it is clear that \td \ depends on \Rmax, $M_1$, and \ah \ . The primary mass $M_1$ determines the white dwarf radius $R_1$ through the $M-R$ relation \citep{Nau72}. Both $M_1$ and \Rmax \ define the disc's extent since the model assumes that the inner disc radius $R_{\rm in}=R_1$.

To compare the observed outbursts decay rates $\taud^o$ with the analytical decay rates $\taud^t$, the derived \td \ has to be divided by the amplitude of the outburst. 
For each system with measured \Porb \ and estimated $M_1$ and $M_2$ (necessary for calculating \Rmax), there is only one free parameter left: \ah. Thus the conformity between the observed decay rate $\taud^o$ and $\taud^t$ calculated from Eq. (\ref{eq:tdec}) can be attained by adjusting \ah. We assumed $\ah =0.2$.
The calculation results are summarized in Table \ref{Tab:UGem}. It is seen that the calculated time is close to the observed one in most cases. In a few cases the discrepancy is large. It is not clear if it is due to the imprecision in measuring the decline time, the peculiar nature of the outbursts, or to the non-universal value of $\ah$. After all, since MRI does not give the correct value of this parameter, we do not really know what physical mechanism drives accretion in hot dwarf nova discs, so there is no reason to assume that it is ``generic".

One should keep in mind, however, that the significant impact on \ah \ has the determination accuracy of the observed outbursts amplitudes. An underestimate of \An \ may be the cause of significantly higher \ah \ for TW~Vir, WX~Hyi, and PTF1J0719.

\section{The Kukarkin-Parenago relation}
\label{sect:KP}

The first to suggest a relation between the outburst amplitude \An \ and the outburst recurrence time \Tn \ were \citet{KP34}. However, since their sample contained outbursts of both recurrent and dwarf novae, its reality has been questioned \citep{PG57,PG77,Bath78}.  In his seminal review article, \citet{Smak84b} attributes the correlation to \citet{PG77} and stresses its statistical character. Finally,  analysing  dwarf nova normal-outbursts data, \citet{VP85} concluded that `the amplitudes and average recurrence times of dwarf novae are correlated".
The most recent version of the Kukarkin-Parenago relation (hereafter K-P relation) in \citet{Warner03} takes the form
\be
\An=(0.7\pm0.43)+(1.9\pm0.22)\log{\Tn},
\label{eq:KP}
\ee
where \Tn \ is in days and \An \ in magnitudes. 

Although it has been argued that the K-P relation might represent some global and average properties of DN outbursts
\citep{VP85}, to the best of our knowledge no derivation from the model has been attempted until now.

While the other relations, connecting various quantities characterizing
the binary systems and their outburst light-curves (such as the absolute visual magnitude-at-maximum $\Vmax(\Porb)$ or the $\taud(\Porb)$
relations \citep[see][and Sect. \ref{sect:tdec}]{Warner03} follow directly from the DIM,  derivation of the
K-P relation is not straightforward. Of course as the outburst amplitude is related
to the mass of the disc and the mass of the disc to the accumulation time, a relation of the K-P form should be expected in principle.
For example, \citet{VP85} speculated that the average amount of mass $\Delta M_{\mathrm{quies}}$ transferred during an
average recurrence time \Tn: $\Delta M_{\mathrm{quies}}=\Tn \Mdot_{\mathrm{quies}}$ is constant over
the dwarf nova population, but also warned about possible selection effects and model dependence.
Here, we try to examine what kind $\An(\Tn)$ relation, if any, can be deduced from the DIM in its simplest (and
when possible simplified) form.

For simplicity we assume that, during quiescence, the accumulation rate \Maccum \ of the mass in the disc
is approximately equal to the mass transfer rate from the secondary \Mtr \ (this assumes no truncation of
the inner edge of the disc; no ``leaky" disc) and that the mean accretion rate during outburst is about half the maximum accretion rate during outburst $\langle\Moutb\rangle\approx\frac{1}{2}\Maccmax$. (This follows from the shape of the function $\Macc(t)$ in the model, which is rapidly rising and then approximately exponentially decreasing during the outburst.)

\begin{figure}[]
\includegraphics[scale=0.9]{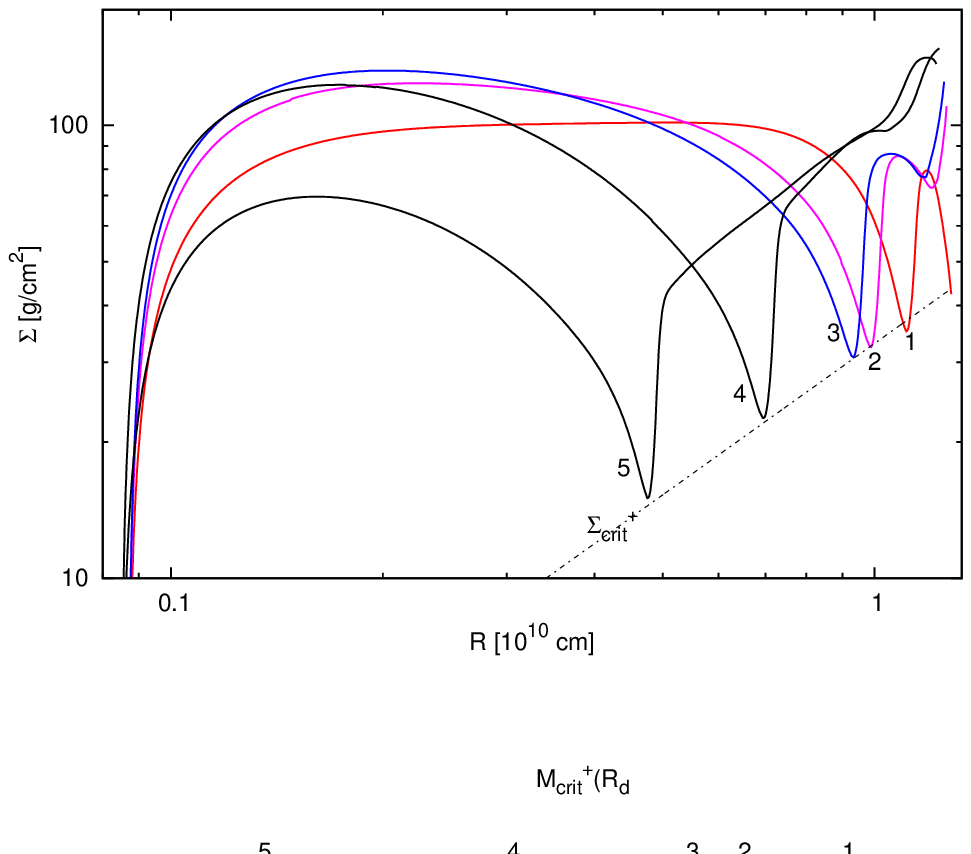}
\includegraphics[scale=0.9]{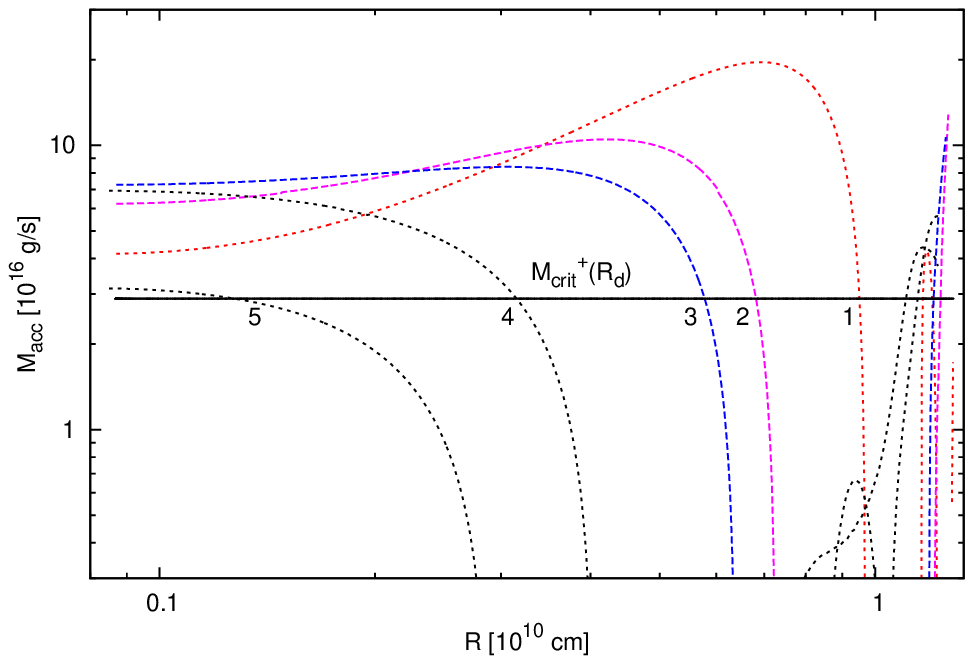}
\caption{\footnotesize \textit{Top}: Evolution of the $\Sigma$ profile during the propagation of the cooling front. Numbers next to the lines stand for the subsequent moments of the front propagation ($1$ is near the moment of the inset of the cooling front). \textit{Bottom}: Evolution of the $\Macc$ profile. The numbers correspond to the same time points as those in $\Sigma$. }
\label{fig:fronty}
\end{figure}

It is worth remarking here that the maximum accretion rate $\Maccmax$  is only approximately equal to $\Mcrp(\Rmax)$ and  simulations clearly show that the mass accretion reaches the maximum only after the cooling front has started to propagate (see Fig. \ref{fig:fronty}).
The accretion rate  keeps rising when the heating front arrives at the outer disc edge and a cooling front starts propagating. This is because just before the launch of the cooling front, the $\Sigma$-profile had not yet reached the stationary hot disc $\Sigma$ shape and the mass from the heated parts near the outer edge had not have time to diffuse fully inwards. During the initial phase of the cooling-front propagation, this mass excess keeps diffusing inwards since it is also ``shoved" by the incoming cooling front. 
The mass accretion rate will eventually drop due to the appearance of the mass shortage in the inner parts of the disc caused by two mechanisms: (1) accretion onto the white dwarf and (2) the strong outflow of mass at the cooling front, which shuffles the mass to the outer parts of the disc (see Sect. \ref{sect:tdec}). 
Guided by the simulation results, we assume $\Maccmax\approx\epsilon\Mcrp(\Rmax)$, with $\epsilon\sim3$ in the following.

The amount of mass accreted during the outburst decay is equal to the mass accumulated
in the disc during quiescence and the rise to outburst maximum: $\Delta M_{\rm accr}=\Delta M_{\rm accum}$. 
\begin{figure}[]
\center
\includegraphics[height=4.2truecm,width=\columnwidth]{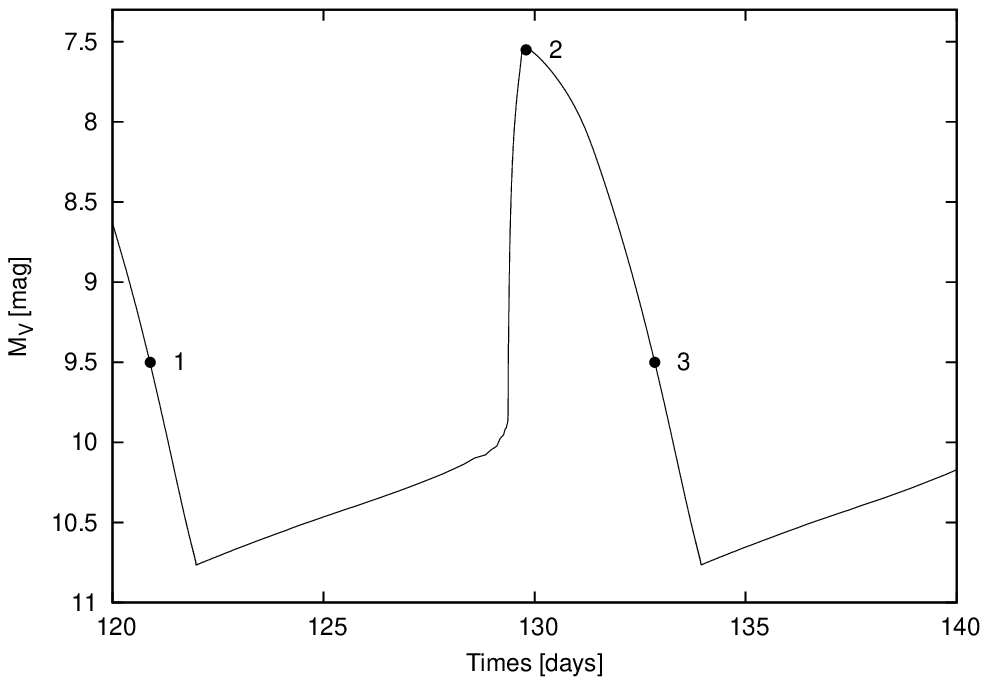}
\includegraphics[height=4.2truecm,width=\columnwidth]{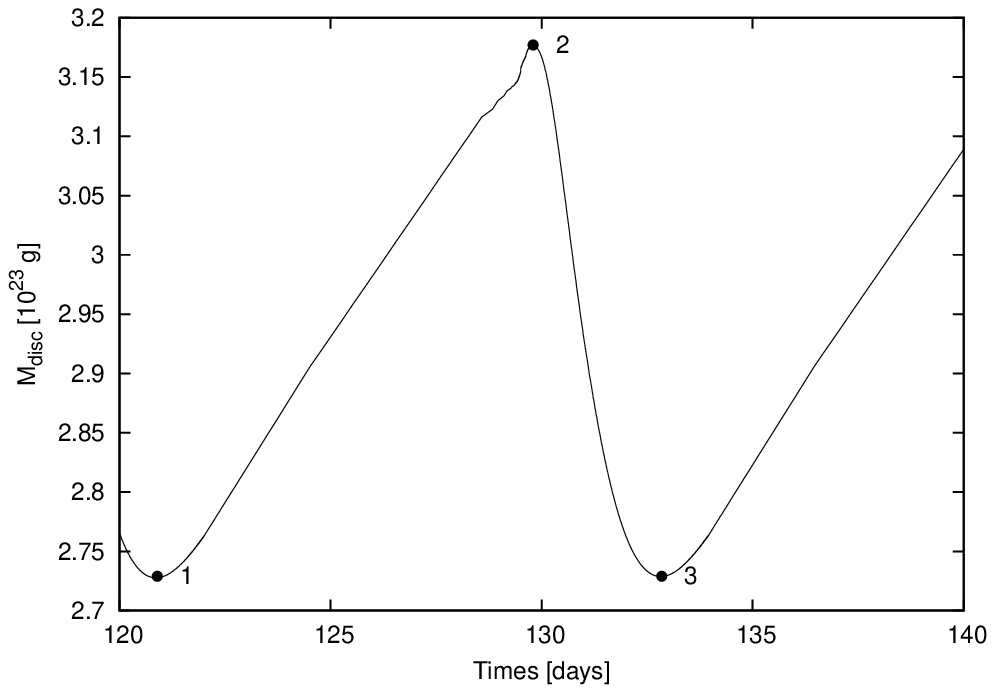}
\includegraphics[height=4.2truecm,width=\columnwidth]{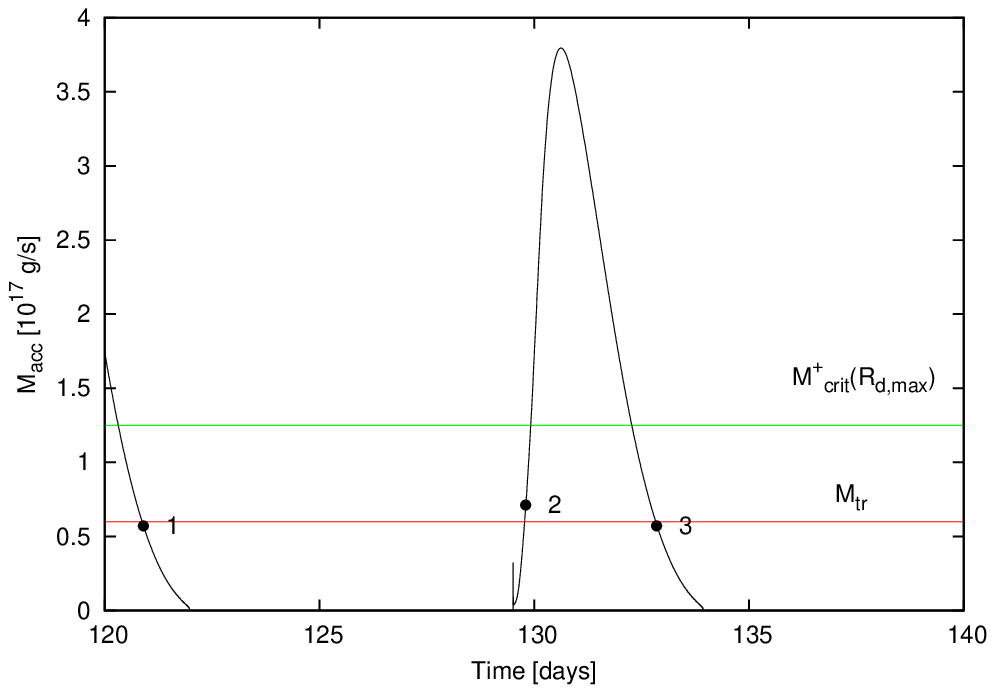}
\caption{{\footnotesize \label{fig:Alfa_KP_cycle}The changes of the disc mass
$M_{\mathrm{disc}}$ (top), magnitude (middle), and mass accretion rate $\Macc$
at $R_{\mathrm{in}}$ (bottom) during one outburst cycle. The horizontal
red line at the $\Macc(t)$ plot is the constant mass transfer rate
$\Mtr$, and the horizontal green line is $\Mcrp(\Rmax)$ for this
model. The parameters of the model are $\ah=0.2$, $\ac=0.05$, $M_{1}=1.0\,\Msun$,
$\Mtr=6.0\times10^{16}\,\,\mathrm{g/s}$, and $\Rave=1.2\times10^{10}\,\,\mathrm{cm}$.
The point number $1$ on all three plots refers to the point where
the mass accumulation in the disc starts, point $2$ is where the
mass depletion from the disc starts, and point $3$ is the end of this
cycle when the mass starts to accumulate in the disc again.}}
\end{figure}

\be
\langle \Moutb \rangle \td=\Mtr(\tq+\tr)
\label{eq:Maccu}
\ee
where \tq \ is the duration of the quiescence, $\tr $ the time it takes the outburst to reach its (bolometric) luminosity maximum, and \td \ the duration of the outburst decay to the quiescence level. Strictly speaking, mass accumulation also occurs during the last part of the decline from maximum (from points 1 to 2 in Fig. \ref{fig:Alfa_KP_cycle}) but this has a negligible effect on the total mass balance.

The decay time \td \ is calculated as described in Section~\ref{sect:tdec}.%
\begin{table}
\caption{Six sets of parameters for which the theoretical K-P relation has been calculated.}
\begin{centering}
\begin{tabular}{|c|c|c|c|c|c|c|c|c|c|c|}
\hline 
 Model & $C_1$ & \ah & \Rmax  & $M_1$  \tabularnewline
\hline 
\hline
$1$ & $-2.5$ &  $0.01$ & $2$ & $1$ \tabularnewline
\hline 
$2$ & $-0.95$ & $0.2$ & $0.6$ & $0.6$   \tabularnewline
\hline 
$3$ & $-0.2$ & $0.2$ & $2$ & $1$ \tabularnewline
\hline 
$4^*$ & $0.1$ & $0.2$ & $0.7$ & $1$ \tabularnewline
\hline 
$5$ & $1.2$ & $0.2$ & $5$ & $1.2$ \tabularnewline
\hline 
$6^*$ & $1.3$ &  $0.2$ & $0.8$ & $1.2$ \tabularnewline
\hline 
\end{tabular}
\par
\end{centering}
\centering{}
\tablefoot{ \ah - hot disc viscosity parameter, and $C_1$ - the constant from Eq. (\ref{eq:An2}) calculated for a given set of parameters. Models $2$ and $5$ correspond to the lower and upper limits of the theoretical K-P relation for solar discs with $\ah=0.2$;  ``$^*$" models are calculated for helium discs ($Y=0.98\,\,Z=0.02$): model $4^*$ gives the K-P relation for PTF1J0719, model $6^*$ gives the K-P relation for CR~Boo and V803~Cen in their cycling states. The models are plotted with lines in Fig. \ref{fig:KPteor}.}
\label{Tab:KP}
\end{table}

The outburst recurrence time $\Tn$ is counted from the onset of the outburst to the onset of the following one ($\Tn=\tr+\td+\tq$), while the accumulation time is $\tq+\tr$.  (\tr \ is non negligible in some type of outbursts, and it is important to include it to be able to account for e.g. the ``cycling state" outbursts where no quiescence phase is present.) Substituting \Tn \ and $t_{\rm accum}$ to Eq. (\ref{eq:Maccu})
gives
\be
\Tn=\left(\frac{\langle \Moutb \rangle}{\Mtr}+1\right)\td=\left(\frac{\Maccmax}{2\Mtr}+1\right)\td
\label{eq:Tn1}
\ee
Since $\Maccmax \approx \epsilon\Mcrp \left(\Rd\right)$ and the
instability condition requires $\Mtr< \Mcrp \left(\Rd\right)$, the ratio in the brackets is $\Mcrp\left(\Rd\right)/(2\Mtr)\gg 1$ and one can take
\be
\Tn \approx \frac{\epsilon\Mcrp\left(\Rd\right)}{2\Mtr}\td ,
\label{eq:Tn2}
\ee
(in what follows we drop the index ``max" in $R_d$).

Assuming that $R_{1}/\Rd\ll 1$, the luminosity at outburst maximum can be approximated as \citep{fkr}
\be
L_{\rm max}\approx \frac{GM_{1}\epsilon\Mcrp\left(\Rd\right)}{2R_1}
\label{eq:Lmax}
\ee
where $R_1$ is a white dwarf radius and $M_1$ its mass.
The luminosity at minimum light can be estimated from the model as
\be
L_{\rm min}=\frac{GM_{1}\Mtr}{2\Rd \tilde{g}}
\label{eq:Lmin}
\ee
where $\tilde{g}\sim 2$ \citep{Idan99}.

The amplitude $\An$ is the difference between the magnitudes at maximum \Vmax \ and  minimum \Vmin \ given by
\be
\An=\Vmin-\Vmax+BC_-^+=2.5\log\frac{L_{\rm max}}{L_{\rm min}}+BC_-^+
\label{eq:An1}
\ee
where $BC_-^+$ is the difference between the bolometric corrections at maximum and minimum.
From Eqs. (\ref{eq:Tn2}), (\ref{eq:Lmax}) and (\ref{eq:Lmin}) we thus get
\be
\An \approx C_{1}+2.5\log \Tn
\label{eq:An2}
\ee
where
\be
C_1=2.5\log\,2\tilde{g}-2.5\log \td +   BC_{\rm max}-BC_{\rm min}.
\label{eq:C1}
\ee
The last step is to estimate the bolometric corrections $BC_{max}$
and $BC_{\rm min}$.

From the definition, $BC_{\rm max}=M_{\rm {bol,max}}-\Vmax$. Following \citet{Smak89} we calculate the visual luminosity at maximum
$L_{\rm{V,max}}$ with spectral energy distribution of a black body disc
integrated over the visual band frequencies. The luminosity $L_{\rm{V,max}}$ (and so $BC_{\rm max}$)
depends on $M_1$, $R_1$, \Rd \ and $\Maccmax$. The bolometric correction $BC_{\rm min}$ can be estimated from the spectral energy distribution calculated for quiescent disc models with effective temperature $\lta 5000\, {\rm K}$. 
We used spectra calculated by Irit Idan \citep[private communication; see][]{ilhs}. Based on that we assumed $BC_{\rm min}\approx-0.4$.

\begin{figure}[]
\includegraphics[scale=0.9]{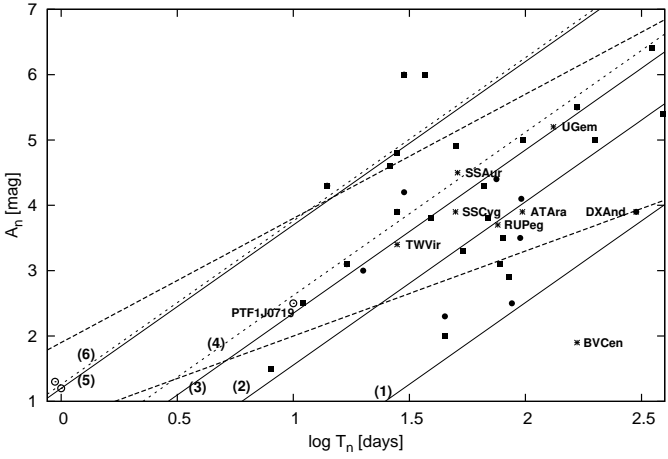}
\caption{\footnotesize Various  $\An(\Tn)$ relations. (Numbers in brackets above the lines correspond to the row numbers in Table \ref{Tab:KP}). The solid lines $(2)$ and $(5)$ correspond respectively to the ``lower and upper limits" (see text) deduced from the theoretical relation for solar-aboundance discs with $\ah=0.2$ ($C_1= -0.95\, \mathrm{and}\, 1.2$). Helium disc theoretical relations are represented by the dotted lines $(4)$ and $(6)$ ($C_1= 0.1\, \mathrm{and}\, 1.3$). Solid lines 1 and 3 illustrate the $\ah$ dependence of  the $\An(\Tn)$ relation: for the same binary parameters, they correspond to $\ah=0.01$ and $\ah=0.2$, respectively. The upper and lower uncertainty of observational K-P relation fitted to the systems are marked with thick, dashed lines. The sample of binaries marked on the plot consists of U~Gem-type systems listed in Table \ref{Tab:UGem} (\textit{full circles and asteriks}), AM CVn systems: PTF1J0719, \object{CR Boo} and \object{V803 Cen} (both in cycling state outbursts) (\textit{open circles}) and SU~UMa stars (normal outbursts only, \textit{full squares}) from \citet{Ak02} and updated Cataclysmic Binaries Catalog \citep{Ritter03}.} 
\label{fig:KPteor}
\end{figure}

Equation (\ref{eq:An2}) does not correspond exactly to the K-P relation (see Eqs. \ref{eq:KP}, \ref{eq:KPour}). The slope of the theoretical relation is always 2.5 (by construction) as compared with $\sim 2$ obtained from fits to observations. Considering the typical spread of parameters,  $C_1$ is contained between $\sim -1.0$ and $\sim 1.2$ in hydrogen-dominated discs, and  is between $\sim -0.2$ and $1.3$ in helium discs (see Table \ref{Tab:KP}), to be compared with $C_1$ between 0.3 and 2.0 for the K-P relation. Considering the very large scatter of observational data, this can be considered a fairly satisfactory result. This is only true of the theoretical relation obtained assuming $\ah=0.2$. The same relation with $\ah=0.01$ gives a totally unacceptable representation of the $\An(\Tn)$ relation, thus confirming the conclusion of the previous sections that  $\ah\approx 0.2$. (Compare models $1$ and $3$ to see how the K-P relation changes when $\ah$ is decreased from $\ah=0.2$ to $=0.01$ in a solar disc.)
The various $\An(\Tn)$ relations are plotted in Fig. \ref{fig:KPteor}. We also marked on this diagram the values of $\An(\Tn)$ for a subset of dwarf novae and outbursting AM Cn stars.
The sample of the systems presented in Fig. \ref{fig:KPteor} consists of U Gem-type binaries taken from \citet{Ak02} and listed in Table \ref{Tab:UGem}, SU~UMa-type binaries taken from \citet{Ak02} (also listed in Table \ref{Tab:UGem}) and from the updated Cataclysmic Binaries Catalog \citep{Ritter03} and three AM~CVn-type stars for which \An \ and \Tn \ were measured from their light-curves. For CR~Boo and V803~Cen  the measured $\An$ and $\Tn$ relate to the outbursts in the cycling state \citep{Patt2000}.

In our sample $15$ systems out of $43$ are the same as used by \citet{Warner03}. The linear fit to our sample gives
\be
\An=(1.3\pm 0.6)+(1.6\pm 0.3)\log\Tn
\label{eq:KPour}
\ee
Upper and lower uncertainties of this relation are marked in Fig.\ref{fig:KPteor}.

Some of the U~Gem-type binaries in Fig. \ref{fig:KPteor} are marked with asterisks and their names as examples to show how observed systems correspond to the theoretical lines. For the same purpose the name of one of AM CVn stars is shown on the plot.

The independence of the theoretical K-P relation from $\Mtr$ is the consequence of the assumption $\Maccmax\approx\epsilon\Mcrp(\Rmax)$ since $\Mcrp(\Rmax)$  does not depend on \Mtr. 
Systems with larger (more extended) discs and more massive $M_1$ have higher $\An$ for a given $\Tn$ than systems with small discs or less massive primaries (compare lines $(2)-(5)$, $(4)-(6)$). 

Despite the simplifications and approximations assumed in the derivation, the theoretical K-P relation follows the observational data reasonably well.
One concludes that normal dwarf nova outbursts are indeed the results of filling and emptying of an accretion disc, as assumed in the model. The parameter that has the deciding influence on the recurrence time and amplitude of normal outbursts is the disc's extent.

\section{Conclusions}
\label{sect:DC}

As in S99, the main conclusion of the present paper is that  in ionized dwarf-nova accretion discs the viscosity parameter $\ah\approx 0.2$. The same conclusion is presumably also true  for helium-dominated discs in outbursting AM CVn stars, although there the statistics on which it is based are fairly poor. Although there is no evidence that the value of $\ah$ is universal, it can be firmly established that, even if it varies over the cataclysmic variable population, it cannot be as low as 0.01, the value resulting from numerical simulations of the MRI, which is the mechanism that is supposed to drive accretion in hot Keplerian discs. Therefore it is not preposterous to suggest that solving this discrepancy between observations and theory should become the main subject of interest of researchers studying disc accretion mechanisms\footnote{After the submission of the present article, a paper on this subject by \citet{LatPap} has been posted on astro-ph.}.

\acknowledgements{}
We are grateful to the referee Ulf Torkelsson for criticism that helped improve our paper. This work has been supported the Polish MNiSW grants PSP/K/PBP/000392,  N N203 380336, the Polish National Science Center grant UMO-2011/01/B/ST9/05439  and the French Space Agency CNES. IK acknowledges the IAP's kind hospitality.

\begin{appendix}

\section{Critical parameters for solar composition discs.}
\label{App}

The method of obtaining the formulae and the formulae themselves can be found in \citet{LDK}.

For the solar composition hydrogen, helium and metal mass fractions $X=0.7\,\,\,Y=0.28\,\,\,Z=0.02$ one obtains from fits to S-curves
\begin{eqnarray}
\Sigma^+ & = & 39.9~\alpha_{0.1}^{-0.80}~R_{10}^{ 1.11}~m_1^{-0.37} \mathrm{g\,cm^{-2}}\nonumber \\
 \Sigma^-  & = &74.6~\alpha_{0.1}^{-0.83}~R_{10}^{ 1.18}~m_1^{-0.40}\mathrm{g\,cm^{-2}}\nonumber\\
 T_{\rm c}^+    &=&30000 ~\alpha_{0.1}^{-0.18}~R_{10}^{ 0.04}~m_1^{-0.01}\mathrm{K}\nonumber\\
 T_{\rm c}^-    &=&8249~\alpha_{0.1}^{0.14}~R_{10}^{-0.1}~m_1^{ 0.04}\mathrm{K}\\
 T_{\rm eff}^+  &=&6890~R_{10}^{-0.09}~m_1^{ 0.03}\mathrm{K}\nonumber\\
 T_{\rm eff}^-  &=&5210~R_{10}^{-0.1}~m_1^{ 0.04}\mathrm{K}\nonumber\\
 \dot{M}^+ &=&8.07\times 10^{15}~\alpha_{0.1}^{-0.01}~R_{10}^{ 2.64}~m_1^{-0.89}\mathrm{g\,s^{-1}}\nonumber\\
 \dot{M}^- &=&2.64\times 10^{15}~\alpha_{0.1}^{0.01}~R_{10}^{ 2.58}~m_1^{-0.85}\mathrm{g\,s^{-1}},\nonumber 
 \label{hecrit}
 \end{eqnarray}
where $\Sigma^\pm$ are the critical surface densities for the hot (+) and cold ($-$) thermal-equilibrium solutions.  Similarly, $T_{\rm c}^\pm$, $T_{\rm eff}^\pm$ are the critical values of mid-plane and effective temperatures, while $\dot{M}\pm$ correspond to critical accretion rates, $m_1$ is the primary mass in solar units and $R_{10}$ is the radius in units of $10^{10}$ cm.

\end{appendix}


\begin{thebibliography}{}
\bibitem[Ak et 
al.(2002)]{Ak02} Ak, T., Ozkan, M.~T., \& Mattei, J.~A.\ 2002, \aap, 389, 478 

\bibitem[Bailey(1975)]{Bailey75} Bailey, J.\ 1975, Journal of 
the British Astronomical Association, 86, 30 

\bibitem[Balbus 
\& Hawley(1998)]{Balbus98} Balbus, S.~A., \& Hawley, J.~F.\ 1998, American Institute of Physics Conference Series, 431, 79 

\bibitem[Bath 
\& Shaviv(1978)]{Bath78} Bath, G.~T., \& Shaviv, G.\ 1978, \mnras, 183, 515 

\bibitem[Dubus et 
al.(2001)]{Dubus01} Dubus, G., Hameury, J.-M., \& Lasota, J.-P.\ 2001, \aap, 373, 251 

\bibitem[Eggleton(1983)]{egg83} Eggleton, P.~P.\ 1983, \apj, 
268, 368 

\bibitem[Frank et al.(2002)]{fkr} Frank, J., King, A., 
\& Raine, D.~J.\ 2002, Accretion Power in Astrophysics, 
Cambridge University Press.

\bibitem[Gilliland(1982)]{BVCen} Gilliland, R.~L.\ 1982, 
\apj, 263, 302 

\bibitem[Hameury et al.(1997)]{H97} Hameury, J.-M., Lasota, 
J.-P., \& Hur\'e, J.-M.\ 1997, \mnras, 287, 937 

\bibitem[Hameury et al.(1998)]{H98} Hameury, J.-M., Menou,
K., Dubus, G., Lasota, J.-P., \& Hur\'e, J.-M.\ 1998, \mnras, 298, 1048



\bibitem[Hirose et al.(2009)]{Hirose09} Hirose, S., Krolik, 
J.~H., \& Blaes, O.\ 2009, \apj, 691, 16 

\bibitem[Idan et al.(1999)]{Idan99} Idan, I., Lasota, J.~P., 
Hameury, J.-M., \& Shaviv, G.\ 1999, \physrep, 311, 213 

\bibitem[Idan et al.(2010)]{ilhs} Idan, I., Lasota, J.-P., Hameury, J.-M., \& Shaviv, G.\ 2010, \aap, 519, A117

\bibitem[King et al.(2007)]{kingetal07} King, A.~R., Pringle, 
J.~E., \& Livio, M.\ 2007, \mnras, 376, 1740 

\bibitem[Kotko et al.(2012)]{KLDH1} Kotko, I., Lasota, J.-P., Dubus, G., \& Hameury, J.-M \ 2012, \aap, 544, A13

\bibitem[Kukarkin 
\& Parenago(1934)]{KP34} Kukarkin, B.~W. \& Parenago, P.~P., \ 1934 Var. Star. Bull. 4, 44

\bibitem[Lasota(2001)]{L01} Lasota, J.-P.\ 2001, \nar, 45,
449

\bibitem[Lasota et
al.(2008)]{LDK} Lasota, J.-P., Dubus, G., \& Kruk, K.\ 2008, \aap, 486, 523

\bibitem[Latter 
\& Papaloizou(2012)]{LatPap} Latter, H.~N., \& Papaloizou, J.~C.~B.\ 2012, arXiv:1207.4727 

\bibitem[Levitan et al.(2011)]{Levitan11} Levitan, D., Fulton, 
B.~J., Groot, P.~J., et al.\ 2011, \apj, 739, 68 

\bibitem[Mason et al.(1988)]{Mason88} Mason, K.~O., Cordova, 
F.~A., Watson, M.~G., \& King, A.~R.\ 1988, \mnras, 232, 779 

\bibitem[Menou et al.(1999)]{Menou99} Menou, K., Hameury, 
J.-M., \& Stehle, R.\ 1999, \mnras, 305, 79 

\bibitem[Menou et al.(2000)]{Menou00} Menou, K., Hameury, 
J.-M., Lasota, J.-P., \& Narayan, R.\ 2000, \mnras, 314, 498 

\bibitem[Meyer 
\& Meyer-Hofmeister(1984)]{MMH84} Meyer, F., \& Meyer-Hofmeister, E.\ 1984, \aap, 132, 143 

\bibitem[Nauenberg(1972)]{Nau72} Nauenberg, M.\ 1972, \apj, 
175, 417

\bibitem[Patterson et al.(2000)]{Patt2000} Patterson, J., 
Walker, S., Kemp, J., O'Donoghue, D., Bos, M., 
\& Stubbings, R.\ 2000, \pasp, 112, 625 

\bibitem[Payne-Gaposchkin(1957)]{PG57} Payne-Gaposchkin, C.\
1957, The Galactic Novae, North-Holland Publishing Company and Interscience Publishers, Chapter 8

\bibitem[Payne-Gaposchkin(1977)]{PG77} Payne-Gaposchkin, C.\
1977, Novae and Related Stars, M. Friedjung ed., Dordrecht, D. Reidel Publishing Co. Astrophysics and Space Science Library, 65, 3

\bibitem[Ritter 
\& Kolb(2003)]{Ritter03} Ritter, H., \& Kolb, U.\ 2003, \aap, 404, 301 


\bibitem[Shakura 
\& Sunyaev(1973)]{SS73} Shakura, N.~I., \& Sunyaev, R.~A.\ 1973, \aap, 24, 337

 \bibitem[Smak(1984a)]{Smak84a} Smak, J.\ 1984, \actaa, 34, 161

 \bibitem[Smak(1984b)]{Smak84b} Smak, J.\ 1984, \pasp, 96, 5 

 \bibitem[Smak(1989)]{Smak89} Smak, J.\ 1989, \actaa, 39, 201 

 \bibitem[Smak(1999)]{Smak99} Smak, J.\ 1999, \actaa, 49, 391, (S99)

 \bibitem[Smak(2001)]{Smak01} Smak, J.\ 2001, \actaa, 51, 
279 

\bibitem[Smak(2005)]{Smak05} Smak, J.\ 2005, \actaa, 55, 315

 \bibitem[Smak(2008)]{Smak08} Smak, J.\ 2008, \actaa, 58, 55 

 \bibitem[Smak(2009a)]{Smak09a} Smak, J.\ 2009a, \actaa, 59, 89 

 \bibitem[Smak(2009b)]{Smak09b} Smak, J.\ 2009b, \actaa, 59, 103

 \bibitem[Smak(2009c)]{Smak09c} Smak, J.\ 2009c, \actaa, 59, 109

 \bibitem[Smak(2009d)]{Smak09d} Smak, J.\ 2009d, \actaa, 59, 121

\bibitem[Smak 
\& Waagen(2004)]{SmakWaagen} Smak, J., \& Waagen, E.~O.\ 2004, \actaa, 54, 433 


\bibitem[Sorathia et al.(2012)]{sorathiaetal12} Sorathia, K.~A., 
Reynolds, C.~S., Stone, J.~M., \& Beckwith, K.\ 2012, \apj, 749, 189 


\bibitem[van 
Paradijs(1983)]{VP83} van Paradijs, J.\ 1983, \aap, 125, L16 

\bibitem[van Paradijs(1985)]{VP85} van Paradijs, J.\ 1985, \aap, 144, 199



\bibitem[Warner(2003)]{Warner03} Warner, B.\ 2003, Cataclysmic Variable Stars, 
(Cambridge Astrophysics Series, 28)

\end{thebibliography}
\end{document}